# Large scale biomedical texts classification: a kNN and an ESA-based approaches

Khadim Dramé[1]*, Fleur Mougin[1], Gayo Diallo[1]

[1] University of Bordeaux, ERIAS, Centre INSERM U897, F-33000 Bordeaux, France

*Corresponding author

Email addresses:

    KD:khadim.drame@u-bordeaux.fr

    FM:fleur.mougin@u-bordeaux.fr

    GD:gayo.diallo@u-bordeaux.fr

# Abstract

**Background**

With the large and increasing volume of textual data, automated methods for identifying significant topics to classify textual documents have received a growing interest. While many efforts have been made in this direction, it still remains a real challenge. Moreover, the issue is even more complexasfull textsare not always freely available. Then, using only partial information to annotatethese documentsis promising but remains a very ambitious issue.

**Methods**

We propose two classification methods: a k-nearest neighbours (kNN)-based approach and an explicit semantic analysis (ESA)-based approach.Although the kNN-based approach is widely used in text classification, it needs to be improved to perform well in this specific classification problem which deals with partial information. Compared to existing kNN-based methods, our method uses classical Machine Learning (ML) algorithms for ranking the labels. Additional features are also investigated in order to improve the classifiers' performance. In addition, the combination of several learning algorithmswithvarious techniques for fixing the number of relevant topics is performed. On the other hand, ESA seems promising for this classification task as it yielded interesting results in related issues, suchas semantic relatedness computation between texts and text classification. Unlike existing works, which use ESA for enriching the bag-of-words approach with additional knowledge-basedfeatures, ourESA-based method builds astandalone classifier. Furthermore, we investigate if the results of this method could be useful as a complementary feature of our kNN-based approach.

**Results**

Experimental evaluations performed on large standard annotated datasets, provided by the BioASQ organizers, show that the kNN-based method with the Random Forest learning algorithm achieves good performances compared with the current state-of-the-art methods, reaching a competitive f-measure of 0.55% while the ESA-based approach surprisingly yielded reserved results.

**Conclusions**

We have proposed simple classification methods suitable to annotatetextual documents using only partial information. They are therefore adequate for large multi-label classification and particularly in the biomedical domain. Thus, our work contributes to the extraction of relevant information from unstructured documents in order to facilitate their automated processing. Consequently, it could be used for various purposes, including document indexing, information retrieval, etc.

## Introduction

The amount of textual data is rapidly growing with an abundant production of digital documents, particularly in the biomedical domain (biomedical scientific articles, medical reports, patient discharge summaries, etc.). Furthermore, thesedataare generally expressed in an unstructured form (i.e., in natural language), which makes its automated processing increasingly difficult. Thus, an efficient access to useful information is challenging. To do so, a suitable representation of textual documents is crucial. Controlled and structured vocabularies, such as the Medical Subject Heading (MeSH®) thesaurus, are widely used to index biomedical texts [1] and consequently to facilitate access to useful information [2][3]. As regards conceptual indexing, concepts defined in thesauri or ontologies are often used to annotate documents. For example, the MEDLINE® citations are manually indexed by the National Library of Medicine® (NLM) indexers using the MeSH descriptors. Although the task of annotators is now facilitated by a semi-automatic method [4], the rapid growth of biomedical literature makes manual-based indexing approaches complex, time-consuming and error-prone [5]. Thus, fully automated indexing approaches seem to be essential. While many efforts have been made to this end, indexing full biomedical texts according to specific segments of these texts, such as their title and abstract, remains a real challenge [6].Furthermore, with the large amounts of data, using only partial information to annotate documents is promising(reduction of computational cost).

In this paper, we propose two classification methods for discovering and selecting relevant topics of new (unannotated)documents: a) a kNN-based approach and b) an ESA-based approach.  Our main contribution is to be able to suggest relevant topics to any new document based solely on portion of it thanks to a classification model learnt from a large collection containing several hundreds of thousands of previouslyannotated documents.

Text classification is the process of assigning labels (categories) to unseen documents. The principle of the kNN-based approach is to consider the set of topics (MeSH descriptors, in this case) assigned manually to the k most similar documents of the target document. Then, these topics are ordered by their relevance score so that the most relevant ones are used to classify the document. In a previous work [5], authors noted that over 85% of MeSH descriptors relevant for classifying a given document are contained in its 20 nearest neighbours. This appearsto better represent the documents rather than what can be found in their title and abstract solely.

First, we have developed a method based on the vector space model (VSM)[7] to determine similar documents. The latter uses the TF.IDF (term frequency – inverse document frequency) weighting scheme for representing documents by vectors constituted by unigrams they contain and the cosine measure for retrieving the document neighbours. Then, we have investigated different types of features and several ML algorithms for selecting relevant topics in order to classify a given document.

On the other hand, ESA [8]has yielded good results in related issues such as semantic relatedness computation between texts [8]and even the text classification[9]. For this reason, we propose to explore it using different association measures in the context where only partial information is exploited for classifying a whole document.

Unlike most works in document classification, our approaches use only partial information (titles and abstracts) of documents in order to predict relevant topics for representing their full content. Since the content of documents is not fully exploited, using large datasets for building the classifiers could be useful for capturing more information. Forthis reason, we used classifiers built from large collections of previously annotated documents. This is a very challenging task, which has motivated the recent launch of BioASQ: an international challenge on large-scale biomedical semantic indexing and question answering[1][6].

The rest of the paper is organized as follows. First, related work concerning biomedical document indexing and, more generally, multi-label classification isreviewedin Section 2. Then, the two proposed methods are detailed in Section 3. In

---

[1] http://bioasq.lip6.fr

Section 4, the experiments are shown while the results are described in Section 5 and discussed in Section 6. Conclusion and future work are finally presented in Section 7.

## Related work

The identification of relevant topics from documents in order to des scribe their content is a very important task widely addressed in the literature. In the biomedical domain, the MTI (Medical Text Insdexer) tool [4] is one of the first attempts to index biomedical documents (MEDLINE citations) using controlled vocabularies. To map biomedical text to concepts from the Unified Medical Language System® (UMLS) Metathesaurus - a system that includes and unifiesmore than 160 biomedical terminologies - the MTI tool uses the well-known concept mapper MetaMap [8] and combines its results with the *PubMed Related Citations* algorithm [10]. The combination of these methods results in a list of UMLS concepts which is then filtered and recommended to human experts for indexing citations. Recently, the MTI was extended with various filtering techniques and ML algorithms in order to improve its performance [11]. Ruch has designed a data independent hybrid system using MeSHfor automatically classifying biomedical texts [12]. The first module is based on regular expressions to map texts to concepts while the second is based on a VSM [7] considering the vocabulary concepts as documents and documents as queries. Then, the rankers of the two components are merged to produce a final ranked list of concepts with their corresponding relevance scores. His results showed that this method achieved good performances, comparable to ML-based approaches. One limitation of this system is that it may return MeSH concepts which match partially the text [1].

ML-based approaches are also proposed to deal with such a task. The idea is to learn a model from a training set constituted of already annotated documents and then to use this model to classify new documents. Trieschnigg et al. [1] have presented a comparative study of six systems which aim at classifying medical documents using the MeSH thesaurus. In their experiments, they showed that the kNN-based method outperforms the others, including the MTI and the approach developed by Ruch[12]. In their work, the kNN classifier uses a language model [13] to retrieve documents which are similar to a given document. The relevance of MeSH descriptors is the sum of the retrieval scores of documents annotated by these descriptors among the neighbouring documents. A similar kNN-based approach has been proposed in [5]. A language model is used to retrieve the neighbours of a given document. Then, a learning-to-rank model [14] is used to compute relevance scores and consequently to rank candidate labels[2] collected from these document neighbours. In this work, the number of labels to classify a document is set to 25. Experiments on two small standard datasets (respectively 200 and 1000 documents) showed that it achieves better performances than the MTI tool.

On the other hand, indexing biomedical documents in which each document of the dataset is assigned one or several categories (also called "labels") can be assimilated as a multi-label classification task. Multi-label classification (MLC) is increasingly studied and especially for text classification purposes [15]. Several methods have been developed to deal with this task [16][17]. They can be categorized into two main approaches [15]: the problem transformation approach [18] and the algorithm adaptation approach [17][19]. The problem transformation approach splits up a multi-label learning problem into a set of single-label classification problems whereas the algorithm adaptation approach adjusts learning algorithms to perform MLC.

---

[2] Labels are categories used to classify documents

In MLC, the kNN-based approach is widely used. This approach has been proven efficient for MLC in terms of simplicity, time complexity, computation cost and performance [17]. Zhang and Zhou [19] proposed a ML-KNN (for Multi-Label kNN) method which extends the traditional kNN algorithm and uses the maximum a posteriori principle to determine relevant labels of an unseen instance. For an instance $t$, the ML-KNN identifies its neighbours and estimates respectively the probabilities that $t$ has and has not a label $l$ based on the training set, for each label $l$. Then, it combines these probabilities with the number of neighbours of $t$ having $l$ as a category to compute the confidence score of $l$. Spyromitros *et al.*[17] propose a similar method, named BR-KNN (for Binary Relevance KNN), and two extensions of this method. The proposed approach is an adaptation of the kNN algorithm using a BR method which trains a binary classifier for each label. Confidence scores for each label are computed using the number of neighbours among the k neighbours that include this label. In [20], an experimental comparison of several multi-label learning methods is presented. In this work, different approaches were investigated using various evaluation measures and datasets from different application domains. In their experiments, authors showed that the best performing method is based on the Random Forest classifier [21]. Other recent works address MLC with large number of labels [22]. Indeed, in many applications, the number of labels used to categorize instances is generally very large. For example, in the biomedical domain, the MeSH thesaurus consisting of thousands descriptors (27,149 in the 2014 version) is often used to classify documents. This large number of descriptors can affect the effectiveness and performance of multi-label models. To address this issue, a label selection based on randomized sampling is performed [22].

# Methods

In this section, we present the text classification approaches developed in our work: a kNN-based approach and an ESA-based approach.

### The kNN-based approach: kNN-Classifier

This approach consists of two steps. First, for a given document, represented by a vector of unigrams, its k most similar documents are retrieved. To do so, the TF.IDF weighting scheme is used to determine the weights of different terms in the documents. Then, the cosine similarity between documents is computed. Once the k nearest documents of a target document are retrieved, the set of labels assigned to them are used for training the classifiers (in the training step) or as candidates for classifying the document (in the classification step). Labels, which are the instances here, are first represented by a set of attributes. Thereafter, ML algorithms are used to build models which are then used to rank candidate labels for annotating a given document. For ranking labels, different learning algorithms are explored.

### Nearest neighbours' retrieval

OurkNN-based approach requires a collection of documents previously annotated for the neighbours' retrieval. For a given document, the aim is to retrieveits k mostsimilar documents. To do so, like the *PubMedRelatedCitations* approach [10], we consider that two documents are similar if they address the same topics. The cosine similarity measure, which is commonly used in text classification and information retrieval (IR) with the VSM [7], is chosen for this purpose.The documents are first segmented into sentences and tokens,whilestop words are removed. From these pre-processed texts, all unigrams are extracted and normalized according to a stemming technique [23]. Then, the cosine measure enables to compute similarity between documents, which are represented by vectors of unigrams. Formally, let $C = \{d_1, d_2, ..., d_n\}$, a collection of $n$ documents, $T = \{t_1, t_2, ..., t_m\}$, the set of terms appearing in the documents of the collection and the documents $d_i$ and $d_j$ being represented respectively by the weighted vectors $d_i = (w_1^i, w_2^i, ..., w_m^i)$ and $d_j = (w_1^j, w_2^j, ..., w_m^j)$, their cosine similarity is defined by [12][25]:

$$Sim(d_i, d_j) = \frac{\sum_{k=1}^{m} w_k^i w_k^j}{\sqrt{\sum_k^m (w_k^i)^2} \sqrt{\sum_k^m (w_k^j)^2}}$$

where $w_k^l$, is the weight of the term $t_k$ in the document $d_l$. It is the TF.IDF value of the term.

In order to optimize the search, the documents in the search space are indexed beforehand using the open source IR API *Apache Lucene*[3]. The k-nearest neighbours' retrieval thus becomes an IR problem where the target document is the query to be processed.

**Collection of candidate labels**
For a given document, once its kNN are retrieved, all labels assigned to these documents are gathered in order to constitute a set of candidate labels likely to annotate this document. Since this can be seen as a classification problem, we use ML techniques to rank these candidate labels. Thus, classical classifiers are used to build classification models which are then exploited to determine the relevant labels for annotating any unseen document. For that purpose, candidate labels are used as training instances (in the training step) or instances to be classified (in the classification step).

**Feature extraction**
To determine the relevance of a candidate label, it is represented by a vector of features (also called attributes). In the training step, its class is set to 1 if the label is assigned to the target document and otherwise 0 while in the classification step, the model uses the label features to determine its class. We defined six features based on related works[5][17].

---
[3] http://lucene.apache.org/core/

For each candidate label, the number of neighbour documents to which it is assigned is used as a feature (Feature 1). This value represents an important clue to determine the class of the label. Moreover, in the classical kNN-based approach, it is the only factor used to classify a new instance. In practice, a voting technique is used to assign the instance to the class that is the most commonamong its k nearest neighbours.

For each candidate label, the similarity scores between the document to classify and its nearest neighbours annotated with this candidate label are summed and this sum is another feature (Feature 2). Since the distance between a document and each of its neighbours is not the same, we consider that the relevance of the labels assigned to them for the target document is inversely proportional to this distance. In other words, the closer a document is to the target document, the more its associated labels are likely to be relevant for the latter. In [1], this is the only feature used to determine the relevance scores of candidate labels.

Formally, like in [17], let $L = \{l_j\}, j = 1, \ldots, n$, be the candidate labels set of a new document d, and $V = \{d_i\}, i = 1, \ldots, k$, its k nearest neighbours, the values of these attributes for the label $l_j$ are respectively defined as:

$$f_1(l_j) = \frac{1}{k} \sum_{i=1}^{n} assigned(l_j, d_i)$$

$$f_2(l_j) = \frac{1}{k} \sum_{l_j \in d_i}^{n} sim(d, d_i)$$

where the binary function $assigned(l_j, d_i)$ returns 1 if the label $l_j$ is assigned to the document $d_i$, 0 otherwise; $sim(d, d_i)$ is the similarity score between the documents $d$ and $d_i$ and is computed using the cosine measure.

For each candidate label, we also checked if all the constituent tokens appear in the title and abstract of the document and consider it as the third feature (Feature 3). This binary feature has been chosen because it captures disjoint terms (terms constituted of disjoint words) which are frequent in the biomedical texts.

In addition to these features, we computed two other features using term synonyms. Indeed, for indexing biomedical documents, the MeSH thesaurus is commonly used. The latter is composed of a set of descriptors (also called main headings) organized into a hierarchical structure. Each descriptor includes synonyms and related terms, which are known as its entry terms. Thus, for each label (called descriptor here), we check whether one of its entries appears in the document. If this is the case, the fourth binary feature (Feature 4) is set to 1 and the descriptor frequency in the document is computed as a value corresponding to the fifth feature (Feature 5), otherwise the two features are set to 0.

Finally, another feature (Feature 6) is used to verify whether a candidate label is contained in the document's title. Our assumption is that if a label appears in the title, it is relevant for representing this document.

The relevance of each of these features is estimated using the information gain measure (Table 1). The first two features mainly permit to compute relevance scores of candidate labels.

**Classifier building**
To build the classifiers, a labelled training set consisting of a collection of documents with their manually associated labels is constituted. For each document in the training set, its nearest neighbours and their manually assigned labels are collected. Each label of this collected set is considered as an instance for the training. Thus, for each label, its different features (see the previous section) are computed. Thereafter, labels obtained from neighbours of the different documents of the training set are gathered to form the training data. Then, classifiers are built from this labelled training data. We have tested the following classification algorithms: Naive Bayes (NB) [24], Decision Trees (DT also known asC4.5 in our case[25]), Multilayer Perceptron (MLP) and Random Forest (RF) [26].We chose these classifiers as they have yielded the best performances in our tests.

For the implementation of these classifiers, we use the WEKA[4] (Waikato Environment for Knowledge Analysis) tool, which integrates many ML algorithms [27], including the four ones we have tested.

**Document classification**
Given a document to be classified, the candidate labels collected from its neighbours are represented as the training ones (see the previous section). The trained model is then used to estimate the relevance score of each candidate label. Indeed, the model computes, for each candidate label, its probabilities to be relevant or not. From these probability measures,the relevancescore of each label is derived. Candidate labels are then ranked according to their corresponding scoresand the N top-scoring ones are selected to annotate the document, where N is determined using three different techniques.

**Selection of the optimal value of N**
In order to determine the optimal value of N, we explore three strategies:

a) Initially, N is set as the number of labels having a relevance score greater than or equal to a threshold arbitrarily set to 0.5. This strategy based only on the relevance score of the label regarding the document is inspired by the original kNN algorithm.

---

[4] http://www.cs.waikato.ac.nz/ml/weka/

b) We then set the value of N as the average size (number of labels assigned) of the sets of labels collected from the neighbours. This strategy has been successfully used for extending the kNN-based method proposed in[17].

c) Finally, in the third strategy, we use the method described in [28]. The principle is to compare the relevance scores of successive labels ofa list of candidate labels ranked in descending order for determining the cut off condition enabling to discard the irrelevant or insignificant ones. This strategy is defined by the following formula:

$$\frac{s_{i+1}}{s_i} \geq \frac{i}{i+1+\alpha}$$

where $s_i$ is the relevance score of a label being at position $i$ and $\alpha$a constant whose optimal value is determined empirically.

**The ESA-based approach**

ESA is an approach proposed for representing textual documents in a semantic way [8]. In this method, the documents are represented in a conceptual space constituted of explicit concepts automatically extracted from a given knowledge base[5]. For this, statistical techniques are used to explicitly represent any kind of text (simple words, fragments of text, entire document) by weighted vectors of concepts. In the approach proposed in [8], the titles of Wikipedia articles are defined as concepts. Thus, each concept is represented by a vector consisting of all terms(except stop words) that appear in the corresponding Wikipedia article. The weight of each word of this vector is the association score between thetermand the correspondingconcept. Thesesscores are computed using the *TF.IDF* weighting scheme [29].

---

[5]Wikipedia in most cases

At the end of this step, each concept is represented by a vector of weighted terms.Then, an inverted index, wherein each term is associated with a vector of its related concepts, is created. In this inverted index, the less significant concepts (i.e., concepts with low weight) for a vector are removed. The index is then used to classify unseentextual documents.

The classification process consists of two steps. For a given document, it is first represented by a vector of terms. The concepts corresponding to these terms are then retrieved in the inverted index and merged to constitute a vector of concepts representing the document. The retrieved concepts are finally ranked according to their relevance score in descending order. The most relevant ones are then selected. This process is illustrated by Figure 1.

Formally, let $T$ be a text, $\{t_i\}$ the termsappearing in $T$ and $\langle v_i \rangle$, their respective weights. Let $\langle k_j \rangle$,be the association score between the term$t_i$ and the concept $c_j$ with $c_j \in C, C$ the set of Wikipedia concepts. The weight of the concept $c_j$ for the text $T$ is defined by:

$$W(c_j) = \sum_{w_i \in T} v_i . k_j$$

Our ESA-based approach explores this technique in the specific case where only partial information is considered (i.e., the title and abstract in the case of scientific articles).First, we assume the availability of concepts (generallydefined in semantic resources) to be used for document classification as well as a labeled training set in which each document is assigned a set of concepts. Unlike the original ESA method where each article is associated with a single concept, in our approach, each document in the training set may be assigned one or more concepts (also called labels here).

From the training set, we use statistical techniques to establish associations between labels and terms extracted from the texts. Thus, for each label, the unigrams that are more strongly associated with it are used for its representation. If the concepts are seen as documents, we face with an IR problem where the goal is to retrieve the most relevant documents (concepts) for a given query (a new document). Therefore, the classical IR models can be used to represent documents and queries, but also to compute the relevance of a document with respect to a given query. In this work, the VSM is used to determine the most relevant concepts for annotating the given document. Like in the kNN-based approach, the documents are processed using the following techniques: segmentation into sentences, tokenization, removal of stop words and normalization using the Porter's stemming algorithm [23].

For computing the association scores between a concept $c$ and a term $t$, we experimented the following measures:

- The *TF.ICF* measure (the TF.IDF scheme adapted to concepts) [7]**:**

$$TF.ICF(t,c) = TF(t,c) * \log \frac{N}{n_i}$$

where $N$ is the total number of concepts, $n_i$ the number of concepts associated with $t$. The factor $TF(t,c)$ is the number of occurrences of $t$ in the documents annotated by the concept $c$ and is defined by:

$$TF(t,c) = \sum_{d \in D_c} \frac{freq(t,d)}{|d|}$$

where $freq(t,d)$ is the frequency of $t$ in the document $d$, $|d|$ is the number of words of $d$ and $D_c$ the set of documents annotated by the concept $c$.

- The *Jaccard coefficient* [30]**:**

$$J(t,c) = \frac{cocc(t,t)}{occ(t) + occ(c) - cocc(t,c)}$$

where $cocc(t,c)$ is the number of documents in which the concept $c$ and the term $t$ co-occur $occ(c)$ is the number of documents annotated by the concept $c$ and $occ(t)$ is the number of documents in which the term $t$ appears.

Finally, to estimate the relevance of a concept to annotate a document, we use the following metric. The relevance score of a concept $c$ for a new document $d$ is defined by:

$$Rel(c,d) = \sum_{w \in d} TF.IDF(t,d) * score(t,c)$$

where $score(w,c)$ is the association score between the term t and the concept $c$ and $TF.IDF(t,d)$ is the TF.IDF value of the term in the document $d$.

## Evaluation

In order to assess the effectiveness of our approaches, we performed two different experiments: one in the context of the task 2a of the international BioASQ challenge to which we participated [31] and the second experiment conducted on a derived dataset from the BioASQ challenge, as described below.

### Datasets

The BioASQ organizers, within the 2014 edition, provided a collection of over 4 million documents constituted by only titles and abstracts of articles (called also citations), coming from specific scientific journals for the task 2a of this challenge [6]. These documents, extracted from the MEDLINE database, are annotated by descriptors of the MeSH thesaurus.

In addition, during the challenge, the organizers provided each week PubMed® citations not yet annotated which were used as test sets to evaluate the systems participating in the task 2a. Participants were asked to classify these test sets using descriptors of the MeSH thesaurus. The test sets have subsequently been annotated by PubMed® human indexers for evaluating the proposals of the participating systems.

**Experiments**
**First experiment**
For the kNN retrieval, we used a dataset consisting of all articles of this collection published since 2000 (2,268,724 documents). The motivation for this choice is to discard old documents which are not annotated by descriptors recently added to the MeSH thesaurus (the MeSH thesaurus is regularly updated). This dataset is thereafter extended to the entire collection.For training the classifiers we randomly selected 20.000 articles out of those published since 2013; the citations of the training set are discarded from the former dataset. We assume this training set sufficient to capture relevant information for building the classifiers.

Only the kNN-based approach was used for our participation to the challenge.To assess this method, five of the different test sets provided by the challenge organizers were used.

**Second experiment**
For the second experiment, we first extracted all articles published since 2013 (133,770 documents) from the previous dataset provided by the challenge organizers. We then selected randomly 20,000 documents to be used for training the classifiers and one thousand for constituting the test set. The data used to train the classifiers werethen extended to 50,000 documents, sincewe believed it could improve the classification performances; using large training dataset should enable the classifiers to capture more information. The test collection was also increased to 2,000 documents. Like in the training dataset, each document in the test set was assigned a set of labels by PubMed® annotators. These manually assigned labels were thus used to evaluate the results of our different methods.

Regarding the evaluation of our ESA-based approach, except the documents in the test set, the rest of the collection (i.e., 4,430,399 documents) was exploited to compute the association scores between words and labels.

**Evaluation measures**

As formerlysaid, indexing biomedical documents can be assimilated to a multi-label classification (MLC) problem. Instead of one class label, each document is assigned a list of labels. Thus, measures usually used for evaluating indexing methods were adapted for the MLC context[15]. The example-based precision (EBP) measures how many of the predicted labels are correct while the example-based recall (EBR) measures how many of the manually assigned labels are retrieved. Since EBP and EBR evaluate partially the performance of a method, the example-based f-measure (EBF) combines both measures for a global evaluation. The accuracy (Acc) is also a complementary measure[15]. These measures are computed as follows. Let $Yi$ be, the set of true labels (labels manually assigned to the documents), $Zi$ the set of predicted labels and $m$ the size of the test set:

$$EBP = \frac{1}{m} \sum_{i=1}^{m} \frac{|Yi \cap Zi|}{|Zi|}$$

$$EBR = \frac{1}{m} \sum_{i=1}^{m} \frac{|Yi \cap Zi|}{|Yi|}$$

$$EBF = \frac{1}{m} \sum_{i=1}^{m} \frac{|Yi \cap Zi|}{|Zi| + |Yi|}$$

$$Acc = \frac{1}{m} \sum_{i=1}^{m} \frac{|Yi \cap Zi|}{|Zi \cup Yi|}$$

These measures, in addition to being common, are representative and enable the global evaluation of the systems' performances.. The results of our two approaches are presented in the next section.

**Experiment environment**

In our different experiments, we used the computing facility of the Bordeaux Mésocentre, Avakas[6], which includes:

- the compute nodes c6100 (x264), which are the machines on which algorithms are executed. They have the following characteristics:
    o Two processors of hexa-cores (12 cores per node) Intel Xeon X5675@3.06GHz;
    o 48GB RAM.
- the computation nodes bigmem R910 (x4), which have more memory and whose cores have slower processors:
    o 4 processors of 10 cores (40 cores per node) Intel Xeon E7-4870@2.4GHz;
    o 512 GB RAM.

In our case, we used two computation nodes c6100, which provide 48 GB of RAM and 24 cores Intel Xeon X5675.

# Results

**Results of the kNN-based approach**
**Experiment within the BioASQ challenge**
First, we present the results obtained in the task 2a of the BioASQ challenge. For that purpose, we report results of batch 3 in terms of EBP, EBR and EBF. We chose only these measures since they are representative and allow estimating the global performance of the MLC methods. Table 2 shows the results of our best system using the kNN-based approach and the ones which obtained the highest measures within the different tests of batch 3. In tests 2 and 5, our best system uses a Naïve Bayes classifier and selects only labels having a confidence score greater than or equal to 0.5 while in the others, the best system sets N to the average size of the sets of labels collected from the neighbours. In most cases, using this value for N yields better or similar results than the other strategy. In the challenge, we do not use the automatic cut-off method to fix the number of labels as described in [28] but in the second experiment, this technique is explored.

---
[6] http://www.mcia.univ-bordeaux.fr/index.php?id=45

**Second experiment**

We evaluate our kNN-based approach with different configurations in the test set of the second experiment and compare the achieved performances. Thus, we test combinations of various classifiers with different techniques for determining the number of labels for annotating a given document. The evaluation of configurations with the two best classifiers in our experiments, NB and RF, are presented in Table 3. The parameter k is empirically set to 25 using a cross-validation technique. When the minimal score threshold is used, the precision often increases significantly, mainly with the RF classifier but the recall is lower. Regarding the "average size strategy", it yields a good recall but the precision decreases slightly. In this case, the results of both classifiers are similar but the RF one slightly outperforms the NB classifier. The best results are achieved with the cut-off method which balances both precision and recall, and yields the best F-measure. Except for the minimum threshold technique where the NB classifier results are better, the best F-measure is achieved with the RF classifier. The DT (C4.5 algorithm of Weka) and the Multilayer Perceptron (MLP) classifiers have also been tested but their results are less interesting. The former yieldslowerperformances while the latter performs very slowly and gets results comparable to the RF ones. The MLP classifier requires more CPU and memory during the training process.

When the training set is raised from 20,000 to 50,000, the performances are slightly improved in two test sets (one of 1000 documents and another of 2000). Table 4 presents the results of the different classifiers in this larger training set. The value of $\alpha$ (constant used in the strategy based on label scores comparison - strategy c- for optimizing N) also affects the classification performance. The lower the value of $\alpha$, the higher the precision is but the lower the recall is and vice versa. In these experiments, we set $\alpha$ to 1.6 which yields the best results using cross-validation techniques. Furthermore, we note that when the classifiers are trained on this extended dataset, they yield similar performances but the RF classifier slightly outperforms the others. Table 5 gives an example of labels suggested for classifying the document having the PMID 23044786 (Figure 2) with the kNN-based approach.

In terms of training time, NB, DT and RF classifiers performed similarly with respectively 4, 6 and 9 minutes once data were represented in suitable format for Weka (e.g. ARFF format (Attribute-Relation File Format)). The pre-processing step (retrieval of neighbours and computation of features values) however takes more time (1 hour and 43 minutes). Note that since we have different types (binary and numeric) of attributes, we discretize the latter in nominal attributes. The MLP classifier is, meanwhile, very costly in terms of training time (23 hours).

**Results of the ESA-based approach**

After processing the training set composed of a collection of 4,432,399 documents (titles and abstracts), we obtain 1,630,405 distinct words and 26,631 descriptors assigned to these documents among the 27,149 MeSH descriptors (98.1%). To simplify the computation and optimize the results of the classification, each concept is represented by a vector consisting of 200 terms, which are the most strongly associated with it. Only terms appearing in at least five documents are considered. Our choice is motivated by the will to simplify the scores computation by excluding the less representative terms. Here, since we used test sets already labelled, the number of concepts which are relevant to annotatethe document is known and is used;therefore, EBP and EBR are equivalent; thus we only report the EBF and the accuracy measures.

After evaluating the ESA-based approach, we note, as in previous work, that its performance varies depending on the measure used to estimate the association scores between words and concepts. This behaviour is illustrated in Table 6 where the Jaccard measure yields the best results.

## Discussion

While textualclassification has been widely investigated, few approaches arecurrently able to efficiently handle large collections of documents, in particular when only a portion of the information is available. This is a challenging task, particularly in the biomedical domain.

Our experiments show that our kNN-based approach is promising for biomedical documents classification in the context of a large collection. Our results confirm the findings presented in [1], where among the multiple classification systems, thekNN-based one yielded the best results. If we compare our method with the latter, we use more advanced features to determine the relevance of a candidate label. Indeed, Trieschnigg and his colleagues determine the relevance of a label by summing the retrieval scores of the k neighbour documents that are assigned to the label[1]. In our method, this sum is only considered as one feature among others for determining the confidence scores of labels. While the results of our method do not outperform the extended (and improved) MTI system [11] which is currently used by the NLMcurators, it getspromising results (0.49 against 0.56 of F-measure). A direct comparison with the method proposed in [5] is not simple since the authors used an older collection than the official datasets provided in the BioASQ challenge, which are recent and annotated with descriptors of the recentMeSHthesaurus (2014 version).Similarly to their experiments, when our method is evaluated on 1,000 randomly selected documents, it outperforms this method (0.55 against 0.50 for the F-measure). But a comparison with their recent results in the first challenge of BioASQ [28] where they integrated the MTI outputs, their system performs better than ours (F-measure of 0.56 against 0.49). Compared with the two approaches proposed in [32], one based on the MetaMap tool [33]and another using IR techniques, our method gets better results (0.49 against 0.42 for the F-measure). Our approach outperforms also the hierarchical text categorization approach proposed in [34].

As part of our participation in the challenge, the NB classifier is combined with the average size of labels assigned to the neighbours to determine relevant descriptors for a given document. In the second experiment, we note however that a combination of RF with the cut-off technique proposed in [28] yields better results [35]. A more recent evaluation of our kNN-based approach using a large dataset (50,000 documents) for training the classifiers shows that it provides better performances, comparable to the best methods described in the literature(with an f-measure of 0.55). Moreover, unlike the extendedMTI system[11], we do not use any specific filtering rules. This makes our approach generic and its reuse in other domains straightforward. A comparison of our basic kNN-based system (trained on 20,000 documents, and improved later) tothe performing classification systems[36][37][38], which alsoparticipatedin the 2014 BioASQ challenge[31] and the baseline (extended MTI)[11]is shown in Table 7.The two best systems, Antinomyra [36] and L2R [38], rely on the learning to rank (LTR) method. The former extends features generated from the neighbour document retrieval with binary classifiers and the results of the MTI and uses then the LTR method to rank the candidate labels. Meanwhile, the other system, combines information obtained by the neighbours' retrieval, binary classifiers and the MTI results as features and also uses theLTRfor the ranking. The Hippocrates system presented in [37]only relies on binary SVM (Support Vector Machine) classifiers and trains them on a large dataset (1.5 million documents) in contrast to our basic kNN approach trained on 20,000 documents. Note that these three systems use binary classifiers for building a model for each label [31]. These systemsrequire therefore considerable resources in terms of computation and storage compared to our kNN-based approach.

For the kNN retrieval, we have investigated the cosine similarity which is widely used in IR. It would be interesting to combine this measure with domain knowledge resources, such as ontologies,to overcome the limitation of similarity computation based only on common words.

The second method based on the ESA, meanwhile, yields very low performances comparable to basic methods using a simple correspondence between the text and the semantic resource inputs. Thus, although the ESA technique has shown interesting results in text classification[9], it does not seem appropriate for our targeted classification problem where only partial information is available. Indeed, to compute the association scores between a term and a label, this method exploits the occurrences of thisterm in the documents annotated by the label. However, in this specific classification problem, labels used to annotate a document are not always explicitly mentioned in the later. Documents are short and it is thereby unlikely that they contain mentions of all relevant labels.It is worth mentioning that in our approach, each concept is represented by a vector consisting of 200 terms, and only terms appearing in at least five documents are considered. For example, the most associated stemmed terms (with their corresponding Jaccard scores) to the label *Body Mass Index* are: index (0.1), waist (0.087), mass (0.079), bodi (0.077), circumfer (0.068), anthropometr (0.062), fat (0.059), adipos (0.048), smoke (0.039), weight (0.038), nutrit (0.037).

Note that we do not use the large Wikipedia's knowledge base,like the work presented in [8], for the conceptual representation of documents since most of the MeSH descriptors cannot be directly mapped to this resource.Furthermore, contrary to existing works [9], which use ESA for enriching the bag-of-words approach with additional knowledge-based features, our ESA-based method builds a standalone classifier. However, this approach will be explored in the futurein order to enrich the features and consequentlyimprove the performance of our k-NN approach.

## Conclusion

In this paper, we have described two approaches for improving the classification of large collections of biomedical documents. The first one is based on the kNN algorithm while the second approach relies on the ESA technique. The former uses the cosine measure with the TF.IDF weighting method to compute similarity between documents and therefore to find the nearest neighbours for a given document. Simple classification methods determine the most relevant labels from a set of candidates of each document. We have investigated an important feature of the classification problem: the decision boundary which permits to determine the relevant label(s) for a target document. Thus, instead of using voting techniques like in the classical kNN algorithm, ML methods were used to classify documents. The latter is based on the ESA technique which exploits associations between words and labels.

Thanks to an evaluation on standard benchmarks, we noted that the kNN based method using the RF classifier with the cut-off method yielded the best results. We also noted that this approach achieved promising performances compared with the best existing methods. In contrast, our findings suggest that the ESA is not suitable for classifying a large collection of documents when only partial information is available. For indexing purpose, the representation of documents as bags of words is limited since similarity between the latter is only based on the words they share. Therefore, to improve the performance of our kNN-based approach, we plan to use a wide biomedical resource, such as the UMLS Metathesaurus, for computing the similarity between documents (exploitation of synonyms and relations) and thus to overcome this limitation. Other features and similarity measures will be studied to improve the performances of our method.

## Competing interests
The authors declare that they have no competing interest.

## Authors' contributions
KD, FM and GD all participated in designing the methods and contributed to the results analysis. KD performed the experiments, discussed the results and drafted the manuscript. GD and FM participated in the correction of the manuscript. All authors read and approved the final version of the manuscript.

## Acknowledgements
The work presented in this paper is supported by the French *Fondation Plan Alzheimer*.The authors would like to thank the BioASQ 2014 challenge organizers who provided the datasets used in this study for evaluating the classification methods. They would also like to thank the anonym reviewers of the previous version of our paper in the *SMBM2014*.

# References

[1] D. Trieschnigg, P. Pezik, V. Lee, F. D. Jong, and D. Rebholz-schuhmann, « MeSH Up: effective MeSH text classification for improved document retrieval », *Bioinformatics*, 2009.

[2] M. C. Díaz-Galiano, M. T. Martín-Valdivia, and L. A. Ureña-López, « Query expansion with a medical ontology to improve a multimodal information retrieval system », *Comput. Biol. Med.*, vol. 39, n$^o$ 4, p. 396‑403, avr. 2009.

[3] M. Crespo Azcárate, J. Mata Vázquez, and M. Maña López, « Improving image retrieval effectiveness via query expansion using MeSH hierarchical structure », *J. Am. Med. Inform. Assoc. JAMIA*, vol. 20, n$^o$ 6, p. 1014‑1020, déc. 2013.

[4] A. R. Aronson, J. G. Mork, C. W. Gay, S. M. Humphrey, and W. J. Rogers, « The NLM Indexing Initiative's Medical Text Indexer », *Stud. Health Technol. Inform.*, vol. 107, n$^o$ Pt 1, p. 268‑272, 2004.

[5] M. Huang, A. Névéol, and Z. Lu, « Recommending MeSH terms for annotating biomedical articles », *J. Am. Med. Inform. Assoc. JAMIA*, vol. 18, n$^o$ 5, p. 660‑667, 2011.

[6] G. Tsatsaronis, M. Schroeder, G. Paliouras, Y. Almirantis, I. Androutsopoulos, É. Gaussier, P. Gallinari, T. Artières, M. R. Alvers, M. Zschunke, and A.-C. N. Ngomo, « BioASQ: A Challenge on Large-Scale Biomedical Semantic Indexing and Question Answering. », in *Information Retrieval and Knowledge Discovery in Biomedical Text, Papers from the 2012 AAAI Fall Symposium, Arlington, Virginia, USA, November 2-4, 2012*, 2012.

[7] G. Salton, A. Wong, and C. S. Yang, « A Vector Space Model for Automatic Indexing », *Commun ACM*, vol. 18, n$^o$ 11, p. 613–620, nov. 1975.

[8] E. Gabrilovich and S. Markovitch, « Computing Semantic Relatedness Using Wikipedia-based Explicit Semantic Analysis », in *Proceedings of the 20th International Joint Conference on Artifical Intelligence*, San Francisco, CA, USA, 2007, p. 1606–1611.

[9] E. Gabrilovich and S. Markovitch, « Wikipedia-based Semantic Interpretation for Natural Language Processing », *J Artif Int Res*, vol. 34, n$^o$ 1, p. 443–498, mars 2009.

[10] J. Lin and W. J. Wilbur, « PubMed related articles: a probabilistic topic-based model for content similarity », *BMC Bioinformatics*, vol. 8, p. 423, 2007.

[11] J. G. Mork, A. Jimeno-Yepes, and A. R. Aronson, « The NLM Medical Text Indexer System for Indexing Biomedical Literature », in *Proceedings of the first Workshop on Bio-Medical Semantic Indexing and Question Answering, a Post-Conference Workshop of Conference and Labs of the Evaluation Forum 2013 (CLEF 2013), Valencia, Spain, September 27th, 2013.*, 2013.

[12] P. Ruch, « Automatic assignment of biomedical categories: toward a generic approach », *Bioinformatics*, nov. 2005.

[13] J. M. Ponte and W. B. Croft, « A Language Modeling Approach to Information Retrieval », in *Proceedings of the 21st Annual International ACM SIGIR Conference on Research and Development in Information Retrieval*, New York, NY, USA, 1998, p. 275–281.

[14] T.-Y. Liu, « Learning to Rank for Information Retrieval », *Found Trends Inf Retr*, vol. 3, n$^o$ 3, p. 225–331, mars 2009.

[15] G. Tsoumakas, I. Katakis, and I. Vlahavas, « Mining multi-label data », in *In Data Mining and Knowledge Discovery Handbook*, 2010, p. 667–685.

[16] E. A. Cherman, M. C. Monard, and J. Metz, « Multi-label Problem Transformation Methods: a Case Study », *CLEI Electron J*, vol. 14, n$^o$ 1, 2011.

[17] E. Spyromitros, G. Tsoumakas, and I. Vlahavas, « An Empirical Study of Lazy Multilabel Classification Algorithms », in *Proceedings of the 5th Hellenic Conference on Artificial Intelligence: Theories, Models and Applications*, Berlin, Heidelberg, 2008, p. 401–406.

[18] J. Read, B. Pfahringer, G. Holmes, and E. Frank, « Classifier Chains for Multi-label Classification », *Mach Learn*, vol. 85, n$^o$ 3, p. 333–359, déc. 2011.

[19] M.-L. Zhang and Z.-H. Zhou, « ML-KNN: A Lazy Learning Approach to Multi-label Learning », *Pattern Recogn*, vol. 40, n$^o$ 7, p. 2038–2048, juill. 2007.

[20] G. Madjarov, D. Kocev, D. Gjorgjevikj, and S. Džeroski, « An extensive experimental comparison of methods for multi-label learning », *Pattern Recognit.*, vol. 45, n$^o$ 9, p. 3084‑3104, sept. 2012.

[21] D. Kocev, C. Vens, J. Struyf, and S. Džeroski, « Ensembles of Multi-Objective Decision Trees », in *Proceedings of the 18th European Conference on Machine Learning*, Berlin, Heidelberg, 2007, p. 624–631.

[22] W. Bi and J. T. Kwok, *Efficient Multi-label Classification with Many Labels.* .

[23] M. F. Porter, « Readings in Information Retrieval », K. Sparck Jones and P. Willett, Éd. San Francisco, CA, USA: Morgan Kaufmann Publishers Inc., 1997, p. 313–316.

[24] G. H. John and P. Langley, « Estimating Continuous Distributions in Bayesian Classifiers », in *Proceedings of the Eleventh Conference on Uncertainty in Artificial Intelligence*, San Francisco, CA, USA, 1995, p. 338–345.

[25] J. R. Quinlan, *C4.5: Programs for Machine Learning*. San Francisco, CA, USA: Morgan Kaufmann Publishers Inc., 1993.

[26] L. Breiman, « Random Forests », *Mach Learn*, vol. 45, n$^o$ 1, p. 5–32, oct. 2001.

[27] M. Hall, E. Frank, G. Holmes, B. Pfahringer, P. Reutemann, and I. H. Witten, « The WEKA Data Mining Software: An Update », *SIGKDD Explor Newsl*, vol. 11, n$^o$ 1, p. 10–18, nov. 2009.

[28] Y. Mao and Z. Lu, « NCBI at the 2013 BioASQ challenge task: Learning to rank for automatic MeSH indexing », Technical report, 2013.

[29] G. Salton and C. Buckley, « Term-weighting Approaches in Automatic Text Retrieval », *Inf Process Manage*, vol. 24, n$^o$ 5, p. 513–523, août 1988.

[30] P. Jaccard, « The Distribution of the Flora in the Alpine Zone », *New Phytol.*, vol. 11, n$^o$ 2, p. 37‑50, 1912.

[31] G. Balikas, I. Partalas, A.-C. N. Ngomo, A. Krithara, and G. Paliouras, « Results of the BioASQ Track of the Question Answering Lab at CLEF 2014 », in *Working Notes for CLEF 2014 Conference, Sheffield, UK, September 15-18, 2014.*, 2014, p. 1181–1193.

[32] D. Zhu, D. Li, B. Carterette, and H. Liu, « An Incremental Approach for MEDLINE MeSH Indexing », in *Proceedings of the first Workshop on Bio-Medical Semantic Indexing and Question Answering, a Post-Conference Workshop of Conference and Labs of the Evaluation Forum 2013 (CLEF 2013), Valencia, Spain, September 27th, 2013.*, 2013.

[33] A. R. Aronson and F.-M. Lang, « An overview of MetaMap: historical perspective and recent advances », *JAMIA*, vol. 17, n$^o$ 3, p. 229–236, 2010.

[34] F. J. Ribadas-Pena, L. M. de C. Ibañez, V. M. D. Bilbao, and A. E. Romero, « Two Hierarchical Text Categorization Approaches for BioASQ Semantic Indexing Challenge », in *Proceedings of the first Workshop on Bio-Medical Semantic Indexing and Question Answering, a Post-Conference Workshop of*

*Conference and Labs of the Evaluation Forum 2013 (CLEF 2013) , Valencia, Spain, September 27th, 2013.*, 2013.

[35] K. Dramé, F. Mougin, and G. Diallo, « A k-nearest neighbor based method for improving large scale biomedical document annotation », in *6th International Symposium on Semantic Mining in Biomedicine (SMBM)*, 2014.

[36] K. Liu, J. Wu, S. Peng, C. Zhai, and S. Zhu, « The Fudan-UIUC Participation in the BioASQ Challenge Task 2a: The Antinomyra system », in *Working Notes for CLEF 2014 Conference, Sheffield, UK, September 15-18, 2014.*, 2014, p. 1311–1318.

[37] Y. Papanikolaou, D. Dimitriadis, G. Tsoumakas, M. Laliotis, N. Markantonatos, and I. P. Vlahavas, « Ensemble Approaches for Large-Scale Multi-Label Classification and Question Answering in Biomedicine », in *Working Notes for CLEF 2014 Conference, Sheffield, UK, September 15-18, 2014.*, 2014, p. 1348–1360.

[38] Y. Mao, C.-H. Wei, and Z. Lu, « NCBI at the 2014 BioASQ Challenge Task: Large-scale Biomedical Semantic Indexing and Question Answering », in *Working Notes for CLEF 2014 Conference, Sheffield, UK, September 15-18, 2014.*, 2014, p. 1319–1327.

## Figures

**Figure 1 - The process of the Explicit Semantic Analysis based approach**
The two steps of the ESA-basedapproach are presented: the indexing step and the classification step.

**Figure 2 - Example of a PubMed® (23044786) citation manually annotated by human indexers using MeSH descriptors**
This is an example of a PubMed citation, consisting of a title and an abstract, with MeSH descriptors manually selected by indexers for annotating it.

## Tables

**Table 1 – Importance of each feature for the prediction according to the Information Gain measure**

| Feature | Description | Information gain |
|---|---|---:|
| Feature 1 | Number of neighbours in which the label is assigned | 0.16 |

| Feature 2 | Sum of similarity scores between the document and all the neighbours' document where the label appears | 0.17 |
| Feature 3 | Check whether all constituted tokens of the label appear in the target document | 0.01 |
| Feature 4 | Check whether one of the label entries appears in the target document | 0.03 |
| Feature 5 | Frequency of the label if it is contained in the document | 0.03 |
| Feature 6 | Check if the label is contained in the document title | 0.02 |

**Table 2 - Results of our kNN-based system and the best systems participating in the BioASQ challenge on the different tests of the batch 3.**

| Test | Number of documents | System | EBP | EBR | EBF |
|---|---|---|---|---|---|
| test 1 | 2,961 | kNN-Classifier | 0.55 | 0.48 | 0.49 |
|  |  | Best | 0.59 | 0.62 | 0.58 |
| test 2 | 5,612 | kNN-Classifier | 0.52 | 0.50 | 0.48 |
|  |  | Best | 0.62 | 0.60 | 0.60 |
| test 3 | 2,698 | kNN-Classifier | 0.55 | 0.49 | 0.49 |
|  |  | Best | 0.64 | 0.63 | 0.62 |
| test 4 | 2,982 | kNN-Classifier | 0.49 | 0.55 | 0.49 |
|  |  | Best | 0.63 | 0.62 | 0.62 |
| test 5 | 2,697 | kNN-Classifier | 0.50 | 0.53 | 0.48 |
|  |  | Best | 0.64 | 0.61 | 0.61 |

**Table 3 - Results of the kNN-Classifier according to the classifier and strategy used for fixing N: a) 0.5 as the minimal confidence score threshold, b) the average size of the sets of labels collected from the neighbours and c) the cut-off method. A training set of 20,000 documents is used.**

| Strategy | Classifier | EBP | EBR | EBF |
|---|---|---|---|---|
| a) | NB | 0.58 | 0.49 | 0.49 |
|  | RF | 0.74 | 0.34 | 0.43 |
| b) | NB | 0.51 | 0.54 | 0.51 |
|  | RF | 0.52 | 0.54 | 0.52 |
| c) | NB | 0.56 | 0.52 | 0.51 |
|  | RF | 0.61 | 0.52 | 0.53 |

**Table 4 - Results of the kNN-Classifier according to the classifier using the cut-off method with a training set of 50,000 documents.**

| Classifier | EBP | EBR | EBF | Acc |
|---|---|---|---|---|
| NB | 0.59 | 0.54 | 0.54 | 0.39 |
| RF | 0.62 | 0.54 | 0.55 | 0.41 |
| C4.5 | 0.63 | 0.52 | 0.54 | 0.39 |
| MLP | 0.64 | 0.46 | 0.51 | 0.36 |

**Table 5 – Labels generated by the kNN-Classifier with their corresponding relevance scores for the document having the 23044786PMID**

| Labels | Relevance | Manual validation |
|---|---|---|
| Humans | 0.99 | Yes |
| Postoperative Care | 0.75 | Yes |
| Female | 0.60 | Yes |
| Male | 0.60 | Yes |
| Middle Aged | 0.32 | Yes |
| General Surgery | 0.32 | Yes |
| Medical Errors | 0.32 | Yes |
| Patient Care Team | 0.32 | No |
| Postoperative Complications | 0.32 | No |
| Adult | 0.26 | Yes |
| Safety Management | 0.26 | No |
| Aged | 0.25 | Yes |
| Prospective Studies | 0.21 | Yes |
| Length of Stay | 0.21 | No |
| Patient Safety | 0.20 | Yes |
| Surgical Procedures, Operative | 0.20 | No |

**Table 6 - Results of the ESA-based approach according to the association score**

| Association score | EBF | Acc |
|---|---|---|
| Jaccard coefficient | 0.26 | 0.16 |
| TF.ICF | 0.22 | 0.13 |

**Table 7 – Comparison of our kNN-Classifierused for participating in the challenge with the best systems and the MTI baseline on the test set of the week 2 of batch 3consisting of 3009 documents. The used measures are: example-based precision (EBP), example-based recall (EBR), example-based f-measure (EBF) and micro f-measure (MiF) (Source BioASQ 2014).**

| Systems | EBP | EBR | EBF | MiF |
| --- | --- | --- | --- | --- |
| Antinomyra [36] | 0.59 | 0.62 | 0.59 | 0.60 |
| L2R [38] | 0.59 | 0.60 | 0.58 | 0.59 |
| Hippocrates [37] | 0.59 | 0.60 | 0.57 | 0.59 |
| MTI | 0.59 | 0.58 | 0.56 | 0.57 |
| kNN-Classifier | 0.55 | 0.49 | 0.49 | 0.51 |